\documentclass[aps,prd,groupedaddress]{revtex4}
\usepackage{graphicx}
\usepackage{amssymb}
\usepackage{epstopdf}
\usepackage{tcolorbox}
\usepackage{color}
\usepackage{float}
\usepackage{enumerate}
\usepackage{ulem}
\usepackage{pdfpages}
\usepackage{amsmath}
\usepackage[colorlinks=true,citecolor=blue,urlcolor=magenta,breaklinks]{hyperref}
\usepackage{graphicx} % Required for inserting images
\usepackage{amsmath}
\usepackage{amssymb}
\usepackage{color}
\usepackage{amsthm}

\theoremstyle{remark}

\usepackage{mathptmx}
\usepackage{tikz}
\usetikzlibrary{arrows.meta}
\usepackage{mathrsfs}

\usepackage{tipa}            % For the thorn symbol
\begin{document}

\title{Traversable ghost wormholes}

\author{Alberto Guilabert}
\email{alberto.guilabert@ua.es}
\affiliation{Fundación Humanismo y Ciencia, Guzmán el Bueno, 66, 28015 Madrid, Spain}
\affiliation{Departamento de F\'{\i}sica, Universidad de Alicante, Campus de San Vicente del Raspeig, E-03690 Alicante, Spain.}

\author{Ernesto Fuenmayor}
\email{ernesto.fuenmayor@ciencs.ucv.ve}
\affiliation{Centro de Física Teórica y Computacional, Escuela de Física, Facultad de Ciencias, Universidad Central de Venezuela, Caracas 1050, Venezuela}

\author{Pedro Bargueño}
\email{pedro.bargueno@ua.es}
\affiliation{Departamento de F\'{\i}sica, Universidad de Alicante, Campus de San Vicente del Raspeig, E-03690 Alicante, Spain.}

\author{E. Contreras }
\email{ernesto.contreras@ua.es }
\affiliation{Departamento de F\'{\i}sica, Universidad de Alicante, Campus de San Vicente del Raspeig, E-03690 Alicante, Spain.}

%%%%%%%%%%%%%%%%%%%%%%%%%%%%%%%%%%
\begin{abstract}
Ghost stars are compact configurations characterized by an arbitrarily small total mass. Such objects require regions of negative energy density—a condition typically regarded as unphysical within the context of conventional stellar models. Nevertheless, negative energy densities arise naturally in traversable wormhole geometries, where the violation of the null energy condition is essential to sustain the flaring-out behavior at the throat. This connection suggests that ghost-like configurations may find a natural realization within wormhole physics. In this work, we investigate the existence of ghost configurations by analyzing their associated Hawking mass. Although in spherical symmetry the Misner and Hawking masses are known to coincide, we show that when the ghost condition is extended beyond spherical symmetry and applied to the Hawking mass, it faces topological obstructions that hinder its straightforward realization. As a concrete example, we demonstrate that a Casimir-like traversable wormhole can be naturally constructed within this framework. Finally, to illustrate the properties of the resulting geometry, we analyze its Penrose–Carter diagram.
\end{abstract}
%%%%%%%%%%%%%%%%%%%%%%%%%%%%%%%%%%%%%%%%%%

\maketitle

Zeldovich and Novikov \cite{Zeldovich1971, Zeldovich1962}, and recently Herrera \cite{herrera2024}, have studied static compact configurations with arbitrarily small total mass. These configurations, termed Ghost Stars in \cite{herrera2024}, are characterized by an energy density that changes sign within the interior, resulting in a vanishing Misner–Sharp mass at the boundary radius. This study focuses on exploring the theoretical feasibility of compact objects with zero total mass (energy) by obtaining exact solutions to Einstein's equations for static spherical distributions of anisotropic fluids, requiring for this the presence of negative energy density in certain regions \cite{LHnegative,Najernegative,Farnesnegative}, which manifests itself indistinctly without a universal distribution pattern. Although the assumptions adopted (conformal flat \cite{herrera2001conformally} or the vanishing complexity\cite{herrera2018new, herrera2018definition, herrera2021complexity, EF3, EF1, arias2022anisotropic, 2020contreras}) have physical meaning, the solutions are offered only to illustrate the possibility of such configurations while a pending challenge is to find exact solutions of ``ghost stars'' directly applicable to astrophysical data. 

Very recent research not only addresses the static solutions that describe their final state, but also seeks to understand their evolution and birth \cite{Herreraevolution, Herrerabirth}, which involves describing the process of gravitational collapse and the intense quantum and radiative effects that lead to this configuration. A particular focus of study is the consideration of ghost stars that require an intrinsic electromagnetic charge to achieve stability or complete their formation \cite{Naseerpossible} (exactly the same aspect has recently been used to raise the possibility of traversable wormholes \cite{turimov2025exact, canate2024traversable,neto2023shadow}). Furthermore, the analysis is expanding to explore the viability of these objects not only under the prevailing spherical symmetry, but also in the context of other spacetime symmetries \cite{Herreraaxially}, seeking a broader understanding of their existence.

In this work we want to emphasize that ghost stars are not compatible with arbitrary space-time geometries and/or topologies, because some topological obstructions may appear when the ghost condition is imposed in arbitrary space-times. Specifically, spherically symmetric and static spacetimes with metric line element given by
\begin{equation}
    ds^2 = -e^{2\phi(r)}dt^2 + \left(1-\frac{2m(r)}{r}\right)^{-1}dr^2 + d\Omega^2,
\end{equation}
which satisfy $2 m/r \leq 1$ everywhere cannot contain any black hole region \cite{Senovilla1998}. Here, $d\Omega^2$ denotes the line element of the 2-sphere.

Although in spherical symmetry the Misner \cite{Misner1964} and Hawking \cite{Hawking1968} masses are known to be equivalent \cite{Hayward1994}, we will see that the ghost condition, namely, a vanishing Misner mass ($m(r_{\Sigma})=0$, with $r_{\Sigma}$ the radius of the surface of the compact object), when considered beyond spherical symmetry and applied to the Hawking mass, leads to some topological obstructions in order to be implemented. In what follows, we examine this issue by first recalling the relevant geometric definitions.

%%
%---------------------------------------------------

Let us consider a spacelike two-surface $\Sigma$. The (Hawking) mass enclosed by $\Sigma$ is defined as
\begin{equation}
    M_{\Sigma}= \frac{1}{2\pi}\sqrt{\frac{A}{16 \pi}}\int _{\Sigma} \star\, \mathrm{Re}\left( K + \rho \rho '\right),
\end{equation}
where: (i) $A$ is the area of $\Sigma$, (ii) $\star$ denotes the Hodge dual on $\Sigma$, (iii) $K$ is a complex curvature commonly referred to as the Penrose–Rindler $K$-curvature~\cite{Penrose1984} and (iv) $\rho$, $\rho'$ are the ingoing and outgoing spin coefficients encoding both expansion and twist of a null geodesic congruence. After using the Gauss–Bonnet theorem, $\int_{\Sigma}\star \, (K + \bar K) = 4 \pi (1-g)$~\cite{Bargueno2023}, where $g$ is the genus of $\Sigma$, we can write
\begin{equation}
    M_{\Sigma}= \frac{1}{2\pi}\sqrt{\frac{A}{16 \pi}}\left(2 \pi \, (1-g)+\int_{\Sigma} \star \, \mathrm{Re}\left(\rho \rho'\right) \right).
\end{equation}
This decomposition reveals an explicit contribution from the topological genus $g$ of the 2-surface $\Sigma$ to the mass formula. Applying the ghost condition $M_{\Sigma}=0$ on the spacelike 2-surface $\Sigma$ imposes a specific constraint on the spin-coefficient integral: $\int_{\Sigma} \star \mathrm{Re}(\rho\rho') = 2\pi(g-1)$, illustrating a topological influence on the dynamic fields required for a ghost configuration.

In the context of static, spherically symmetric spacetimes, we have $g=0$. Moreover, the product $\rho\rho'$ is constant over $\Sigma$, which implies that the expansions coefficients have opposite signs, i.e. $\Sigma$ cannot be a trapped surface. This is consistent with the bound $2m(r)/r \le 1$ throughout the interior. We therefore conclude that static, ghost-interior solutions of this type do not contain closed trapped surfaces. We remark that while wormhole throats (which are minimal surfaces, not trapped surfaces) are not precluded by this analysis, trapped surfaces of spherical topology are (in broader mathematical contexts, trapped surfaces with higher genus $g \ge 1$ are known to be possible~\cite{Mena2008}, but they do not occur in the spherically symmetric setting considered in this manuscript).
Then, although these results are compatible with applying the ghost condition to interior solutions as commented before, it is noticeable that wormhole spacetimes, which do not have trapped surfaces, are also allowed. This moves us to the second point of the manuscript.

%--------------------------------------------
%Wormholes were first proposed by Flamm \cite{flamm} in the context of the Schwarzschild solution through isometric embedding. Later, Einstein and Rosen \cite{Einstein} introduced a geometric structure now known as the Einstein-Rosen bridge, which Wheeler later termed a ``wormhole'' \cite{Wheeler, misner}. These tunnel-like geometries connect either two separate universes or distinct regions of a single universe, and have long captivated theoretical physicists, especially within the framework of spherical symmetry. Traversable wormholes, popularized by the seminal work of Morris and Thorne \cite{thorne}, are characterized by the presence of a throat—representing a minimal surface—that links two asymptotic regions. Since then, a vast number of wormhole solutions have been investigated. Among the most studied is the Ellis wormhole, a classic solution within General Relativity \cite{E1,E2}. For recent developments, see \cite{Kim:2025zyo,Das:2025oro,Magalhaes:2025nql,Muniz:2025qyv,Reboucas:2025maq,Tayde:2025aay,Jaiswal:2025gkw,Tayde:2025yzq,Furtado:2025zva, WH-EF} and references therein.

Interestingly, the existence of ghost stars require negative energy density regions, a condition often considered unnatural in the context of ordinary stellar models. However, these negative energy densities arise naturally in the framework of traversable wormholes, where the violation of the null energy condition (NEC) is essential to sustain the flaring-out of the geometry at the throat \cite{wormholes_1988_morris, morris1988wormholes}. In this sense, the consideration of ghost wormholes (adopting the terminology in \cite{herrera2024}), defined as traversable wormholes with vanishing total mass, becomes natural. Now, the NEC can be violated if the fluid has a negative energy density greater in magnitude than its radial pressure. Although negative energy is conceptually non-intuitive, it is not physically implausible. One prominent mechanism proposed to generate it in the laboratory is the Casimir effect, which arises between two uncharged, closely spaced parallel metallic plates in vacuum. The effect was theoretically predicted in 1948 \cite{casimir1948} and experimentally investigated soon after at Philips Laboratories \cite{sparnaay1957,sparnaay1958}, with more precise confirmations emerging in recent decades \cite{lamoreaux1997}. A particularly intriguing feature of the Casimir effect is that it produces an attractive force due to a region of negative energy density.

Traversable wormholes supported by Casimir energy have been considered in \cite{Garattini2019,ford1996,visser1995,morris1988, avalos2022traversable}. In particular, \cite{Garattini2019} claims that a Casimir wormhole can be sustained by an arbitrarily small amount of exotic matter, since the quantifier of exoticity can be made arbitrarily small. However, because the energy density is spread throughout the entire spacetime, this poses a challenge in the context of laboratory-based Casimir energy, which is generated only in confined regions. Therefore, we can only ensure the existence of negative energy in a small, localized region, and not across the entire spacetime. It is worth highlighting the fact that other attempts have been made recently, introducing in addition to the aforementioned effect, conditions on the symmetry of space (hyperbolic, for example) or even characterizing the complexity as a useful tool and well-supported condition for studying such systems \cite{EF3,avalos2022traversable, avalos2025hyperbolic, sokoliuk2022probing, EF2, bhattacharya2023complexity, EFD, ziaie2024casimir, EF0, varsha2025novel,Varsha:2025oik,EF4}.

Here we propose an alternative approach to constructing traversable wormholes confined to a localized region in a natural and consistent manner. Although wormhole geometries can in principle be matched to the Schwarzschild vacuum, such matching typically requires the presence of a thin shell, as the continuity of the second fundamental form is generally not guaranteed. To circumvent this issue, we draw inspiration from the ghost condition. Here, we extend this concept to the wormhole context by constructing traversable geometries with arbitrarily small mass functions, entirely confined within a finite region that can be made as small as desired. The effective mass of such configurations can be made to vanish or remain negligibly small, making them especially relevant for modeling wormholes supported by effects like the Casimir energy, which is known to be extremely weak (of the order of $10^{-24}\,\mathrm{kg}$).

We are interested in studying static and spherically symmetric wormholes between two disconnected flat regions. In what follows we shall briefly describe its main features. First, without loss of generality, the line element can be parametrized using the usual Schwarzschild coordinates as
\begin{eqnarray}\label{eq:metric}
ds^{2}=-e^{2\phi(r)}dt^{2}+\left(1-\frac{b(r)}{r}\right)^{-1}dr^2+r^{2}d\Omega^{2},
\end{eqnarray}
where $\phi(r)$ is called the redshift function and $b(r)$ the shape function. Each patch is covered by the coordinate range $[r_0,+\infty)$, and the throat of the wormhole is located at $r = r_0$. Here, $r_0$ denotes the radius of the throat of the wormhole and is determined by the minimum value of $r(l)$ where $l$ is the proper radial distance. Note that the information on the Misner-Sharp mass is encoded in the shape function. As usual, the throat condition, i.e. $dr/dl = 0$, demands the existence of a minimum length $r_{0}$ where $b(r_{0})=r_{0}$. Second, the Einstein field equations are given by
\begin{eqnarray}
\rho &=& \frac{b'}{8 \pi  r^2}\label{rho}\\
p_{r}&=&\frac{-2 r b \phi '-b+2 r^2 \phi '}{8 \pi  r^3}\label{pr}\\
p_{t}&=&\frac{\left(r \phi '+1\right) \left(-r b'+2 r (r-b) \phi '+b\right)+2 r^2 (r-b) \phi ''}{16 \pi  r^3}
\end{eqnarray}
and, third, the flaring--out condition is

\begin{eqnarray}\label{foc}
\frac{b/r - b' - 2(r-b)\phi'}{|b'|}>0,
\end{eqnarray}
which at the throat of the wormhole, $r=r_{0}$, reads
\begin{equation}
b'(r_0)<1    
\end{equation}
and  leads to the violation of the NEC, namely $\rho + p_{r} < 0$.
%Let us define the quantity
%\begin{eqnarray}
%\xi=-\frac{p_{r}+\rho}{|\rho|}=\frac{b/r-b'-2(r-b)\phi'}{|b'|},
%\end{eqnarray}
%which can be written as
%\begin{eqnarray}
%\xi=\frac{2b^{2}}{r|b'|}\frac{d^{2}r}{dz^{2}}
%-2(r-b)\frac{\phi'}{|b'|}
%\end{eqnarray}
%Now, as $(r-b)\to0$ at the throat, we have 
%\begin{eqnarray}
%\xi=\frac{2b^{2}}{r|b'|}\frac{d^{2}r}{dz^{2}}>0
%\end{eqnarray}
%so that
%\begin{eqnarray}
%\xi=-\frac{p_{r}+\rho}{|\rho|}>0.
%\end{eqnarray}
Note that if $\rho < 0$, the above condition implies either $p_r < 0$ (in which case $T^1_{\ 1}$ should be interpreted as a tension), or $p_r > 0$, in which case one must ensure that $|\rho| + p_r > 0$ in order for the violation of the energy condition to persist.

As briefly commented before, we define a static and spherically symmetric ghost wormhole as a traversable wormhole defined in the interval $r\in[r_0,r_\Sigma]$ such that $b(r_\Sigma)=0$. We note that, in absence of radial tidal forces, which implies $\phi=0$, the NEC is violated in the whole wormhole geometry (not only near the throat) ) if $b(r)/r$ is a monotonously decreasing function\footnote{Even more, the flare out condition also imposes $r^{-1}(b/r)'=(b'-b/r)<0$ near the throat, which is of the same sign as $\rho+p_r$. Thus, the NEC must be violated near the throat.}. This result can be proven directly combining  (\ref{rho}) and (\ref{pr}) with $\phi=0$ we obtain
\begin{eqnarray}
\rho+p_{r}=\frac{1}{8\pi r}\frac{d}{dr}\left(\frac{b(r)}{r}\right)    
\end{eqnarray}
from which, if $b/r$ is monotonously decreasing, $\rho+p_{r}<0$. As a particular example we propose
\begin{eqnarray}
b(r)=\frac{r_{0}^{2}}{r}\frac{r_\Sigma-r}{r_\Sigma-r_0},
\end{eqnarray}
from which
\begin{eqnarray}
\rho=-\frac{r_{0}^{2} r_\Sigma}{8\pi r^{4}(r_\Sigma-r_0)}\\
p_{r}=-\frac{r_{0}^{2}(r_\Sigma-r)}{8\pi r^{4}(r_\Sigma-r_0)}
\end{eqnarray}
(The expression for the tangential pressure is more involved and not shown here).

At this point a couple of comments are in order. 

First, note that the solution is regular everywhere in the region $r_{0}<r<r_{\Sigma}$. In fact, 
the curvature scalars (the Ricci scalar $R$, the Ricci squared invariant $R_{\mu\nu}R^{\mu\nu}$, and the Kretschmann scalar $\mathcal{K} = R_{\mu\nu\lambda\rho}R^{\mu\nu\lambda\rho}$)  are given by
\begin{eqnarray*}
 R&=&\frac{2 r_{0}^2 r_{\Sigma}}{r^4 (r_{0}-r_{\Sigma})}\\
Ricc&=
&\frac{r_0^4 \left( 3r^2 - 8r\,r_{\Sigma} + 8r_{\Sigma}^2 \right)}
     {2r^8 \left( r_0 - r_{\Sigma} \right)^2}\\
\mathcal{K}&=&     \frac{2 r_0^4 \left( 3r^2 - 8r\,r_{\Sigma} + 6r_{\Sigma}^2 \right)}
     {r^8 \left( r_0 - r_{\Sigma} \right)^2}
\end{eqnarray*}

Second, note that the energy density resembles the Casimir one which is given by
\begin{eqnarray}
    \rho_C(a) = -\frac{\hbar c \pi^2}{720 a^4}
\end{eqnarray}
with $a$ the separation between the plates, in the sense that our solution scales as $r^{-4}$. However, this analogy must be treated with care, since in the Casimir case the energy density is constant, whereas in our case it decreases as a power of $r$. Nonetheless, because our solution has a finite size, namely $r_{0} < r < r_{\Sigma}$, we can choose the separation to be arbitrarily small (while avoiding the limit $r_{0} = r_{\Sigma}$), so that the variation in $r$ becomes negligible and the behaviour approaches the Casimir requirement.
Next, the radial pressure can be written in terms of then energy density as
\begin{eqnarray}
p_{r}=\rho+\alpha(-\rho)^{-1/4}    
\end{eqnarray}
with $\alpha=(r_{0}^{2}r_\Sigma/8\pi(r_\Sigma-r_0))^{-1/4}$ which confirm that the analogy is merely formal, because,
as stated in \cite{Garattini2019}, the equation of state relating the energy density and the radial pressure in a Casimir wormhole is given by $p_r = \omega \rho$, with $\omega = 3$. In contrast, our relation includes not only a different proportionality constant but also an additional term.

The smooth \footnote{The metric tensor should be at least $\mathcal{C}^2$ at some neighbourhood containing $\Sigma$ so field equations are well defined. This smoothness requirement can be relaxed to $\mathcal{C^1}$ if one allows the presence of a thin shell, as commented above. In such a case, the field equations are only satisfied in the distribution sense.} gluing of the wormhole solution, at the hypersurface $\Sigma$ defined by $r=r_{\Sigma}$, with an exterior patch is ensured by the junction conditions introduced by Darmois and Israel \cite{Darmois1927, Israel1966}. Considering an exterior with line element given by Eq. \eqref{eq:metric}, the first junction condition imposes $b(r_{\Sigma}) = 0$ and $\phi(r_{\Sigma}) = 0$. Moreover, the second junction condition implies that $\left[\mathbb{I}_{ab}\right] = 0$, where $\mathbb{I}$ is the second fundamental form induced on the hypersurface\footnote{Note that the continuity on the second fundamental form is, in general, a weaker requirement than imposing continuity of the pressure at $r_{\Sigma}$, see \cite{Bonnor1981}.}, from which we obtain the relation $\phi'(r_{\Sigma}) = 0$. Those conditions require the exterior patches to be Minkowski regions.

As shown in FIG. 1, the resulting spacetime consists of two flat Minkowski regions (\texttt{I} and \texttt{III}) that are smoothly connected through the wormhole interior (region \texttt{II}) across the hypersurface $\Sigma$ (and analogously across $\Sigma'$). Additionally, the diagram shows how null rays can propagate from one asymptotic region to the other through the wormhole throat.

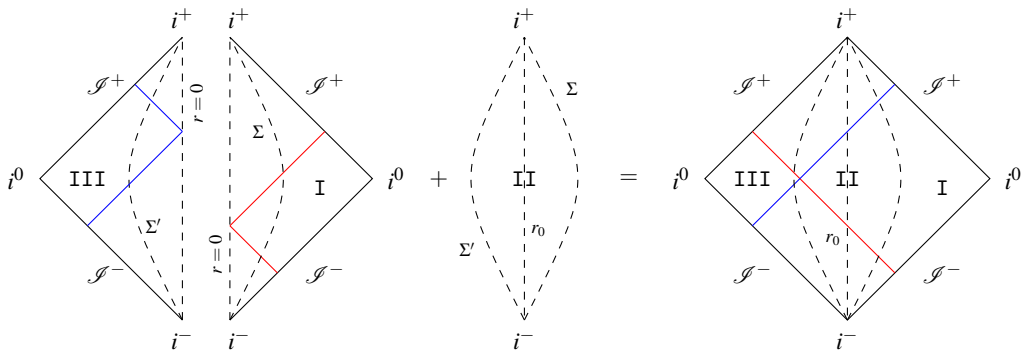
\begin{figure}
\label{fig1}
\centering
\begin{tikzpicture}[scale=1.25]
% Primer diagrama (izquierda)
% Region izq
\draw (-4,1.5) -- (-5.5,0) -- (-4,-1.5);
\draw[dashed] (-4,-1.5) -- (-4,1.5);
\node at (-4.8,1) {$\mathscr{I}^+$};
\node at (-4.8,-1) {$\mathscr{I}^-$};
\node at (-4,1.7) {$i^+$};
\node at (-4,-1.7) {$i^-$};
\node at (-5,0.) {$\texttt{III}$};
\node at (-5.75,0) {$i^0$};
\draw[dashed] (-4,-1.5) .. controls (-4.75,0) .. (-4,1.5);
\draw[blue] (-5,-0.5) -- (-4,0.5);
\draw[blue] (-4,0.5) -- (-4.5,1);
\node at (-4.3,-0.5) {\scriptsize$\Sigma'$};
\node[rotate=90] at (-3.85,0.8) {\scriptsize$r=0$};

% Region der
\draw (-3.5,1.5) -- (-2,0) -- (-3.5,-1.5);
\draw[dashed] (-3.5,-1.5) -- (-3.5,1.5);
\node at (-2.5,1) {$\mathscr{I}^+$};
\node at (-2.5,-1) {$\mathscr{I}^-$};
\node at (-1.75,0) {$i^0$};
\node at (-2.55,-0.1) {$\texttt{I}$};
\node at (-3.2,0.5) {\scriptsize$\Sigma$};
\draw[red] (-3,-1) -- (-3.5,-0.5);
\draw[red] (-3.5,-0.5) -- (-2.5,0.5);
\draw[dashed] (-3.5,-1.5) .. controls (-2.75,0) .. (-3.5,1.5);
\node at (-3.4,1.7) {$i^+$};
\node at (-3.4,-1.7) {$i^-$};
\node[rotate=90] at (-3.65,-0.8) {\scriptsize$r=0$};

% Suma
\node at (-1.3,0) {$+$};

% Segundo diagrama (centro)
\draw[dashed] (-0.4,1.5) -- (-0.4,-1.5);
\draw[dashed] (-0.4,-1.5) .. controls (0.35,0) .. (-0.4,1.5);
\draw[dashed] (-0.4,-1.5) .. controls (-1.15,0) .. (-0.4,1.5);
\node at (-0.4,1.7) {$i^+$};
\node at (-0.4,-1.7) {$i^-$};
\node at (-0.4,0) {$\texttt{II}$};
\node at (-0.25,-0.55) {\scriptsize$r_0$};
\node at (-1,-0.75) {\scriptsize$\Sigma'$};
\node at (0.1,0.95) {\scriptsize$\Sigma$};

% Igualdad
\node at (0.7,0) {$=$};

% Tercer diagrama
\draw[dashed] (3,1.5) -- (3,-1.5);
\draw (3,1.5) -- (4.5,0) -- (3,-1.5) -- (1.5, 0) -- cycle;
\node at (2,0.) {$\texttt{III}$};
\node at (3.,0) {$\texttt{II}$};
\node at (4,-0.1) {$\texttt{I}$};
\draw[red] (3.5,-1) -- (2,0.5);
\draw[blue] (2,-0.5) -- (3.5,1);
\node at (2.85,-0.65) {\scriptsize$r_0$};
\node at (4.75,0) {$i^0$};
\node at (1.25,0) {$i^0$};
\node at (4,1) {$\mathscr{I}^+$};
\node at (4,-1) {$\mathscr{I}^-$};
\node at (2,1) {$\mathscr{I}^+$};
\node at (2,-1) {$\mathscr{I}^-$};
\node at (3,1.7) {$i^+$};
\node at (3,-1.7) {$i^-$};
\draw[dashed] (3,-1.5) .. controls (2.25,0) .. (3,1.5);
\draw[dashed] (3,-1.5) .. controls (3.75,0) .. (3,1.5);
\end{tikzpicture}

\caption{Schematic representation of the ghost wormhole. Here, regions \texttt{I} and $\texttt{III}$ denote disconnected Minkowski patches and region \texttt{II} denotes the wormhole interior. Two light rays are shown to illustrate the traveling through the wormhole interior.}
\end{figure}

To conclude, we briefly summarize. In this work we extend the concept of ghost stars (compact configurations with arbitrarily small total mass) to the framework of traversable wormholes. Ghost stars require regions with negative energy density, a feature typically regarded as unphysical in stellar models but naturally arising in wormhole geometries due to the violation of the null energy condition. We show that imposing the ghost condition on quasilocal mass definitions introduces topological constraints on admissible spacetimes. In spherical symmetry, the Misner--Sharp and Hawking masses coincide, allowing the implementation of ghost conditions consistently. Motivated by these results, we construct static and spherically symmetric ghost wormholes, defined as traversable wormholes with vanishing total mass confined within a finite region. Explicit examples are provided where the effective energy density resembles Casimir-like contributions, supporting the interpretation of ghost wormholes as configurations localized in regions of exotic matter. Our analysis highlights that such solutions can be sustained by extremely small negative energy densities, thereby reinforcing the plausibility of wormholes supported by localized quantum vacuum effects.

\section*{Acknowledgements}
A. G. acknowledges Fundación Humanismo y Ciencia for financial support. E. F. is grateful for  MINCYT-CDCH-UCV/ 2024  and acknowledges support from Consejo de Desarrollo Científico y Humanístico - Universidad Central de Venezuela in part by a grant entitled Study of compact stellar configurations composed of spherically symmetric and hyperbolic, static and anisotropic relativistic fluids in the context of General Relativity. Also, E. F. is grateful for the Funcaci\'on Carolina-$25$ years for granting the short-stay scholarship that served to finance this work. E. F. expresses sincere gratitude to the group of Astrof\'isica Relativista at the Universidad de Alicante for their warm hospitality during the development of this work.
 P. B. and E. C. acknowledge financial support from the Generalitat Valenciana through PROMETEO PROJECT CIPROM/2022/13. E. C. is funded by the Beatriz Galindo contract BG23/00163 (Spain).

\bibliographystyle{unsrt}
\bibliography{main}

@book{Zeldovich1971,
  author = {Ya. B. Zeldovich and I. D. Novikov},
  title = {Relativistic Astrophysics. Vol I. Stars and Relativity},
  publisher = {University of Chicago Press},
  year = {1971},
  address = {Chicago, Illinois}
}

@article{EFD,
  title = {Gravitational cracking of general relativistic polytropes: A generalized scheme},
  author = {Le\'on, P. and Fuenmayor, E. and Contreras, E.},
  journal = {Phys. Rev. D},
  volume = {104},
  issue = {4},
  pages = {044053},
  numpages = {15},
  year = {2021},
  month = {Aug},
  publisher = {American Physical Society},
  doi = {10.1103/PhysRevD.104.044053},
  url = {https://link.aps.org/doi/10.1103/PhysRevD.104.044053}
}

@article{Zeldovich1962,
  author = {Ya. B. Zeldovich},
  title = {The Collapse of a Small Mass in the General Theory of Relativity},
  journal = {Sov. Phys. JETP},
  volume = {15},
  pages = {446},
  year = {1962}
}

@article{Bonnor1981,
  title = {Junction conditions in general relativity},
  volume = {13},
  ISSN = {1572-9532},
  url = {http://dx.doi.org/10.1007/BF00766295},
  DOI = {10.1007/bf00766295},
  number = {1},
  journal = {General Relativity and Gravitation},
  publisher = {Springer Science and Business Media LLC},
  author = {Bonnor,  W. B. and Vickers,  P. A.},
  year = {1981},
  month = jan,
  pages = {29–36}
}

@article{Morris1988,
  title = {Wormholes,  Time Machines,  and the Weak Energy Condition},
  volume = {61},
  ISSN = {0031-9007},
  url = {http://dx.doi.org/10.1103/PhysRevLett.61.1446},
  DOI = {10.1103/physrevlett.61.1446},
  number = {13},
  journal = {Physical Review Letters},
  publisher = {American Physical Society (APS)},
  author = {Morris,  Michael S. and Thorne,  Kip S. and Yurtsever,  Ulvi},
  year = {1988},
  month = sep,
  pages = {1446–1449}
}

@article{morris1988wormholes,
  author = {M. S. Morris and S. Kip Thorne and U. Yurtsever},
  journal = {Physical Review Letters},
  volume = {61},
  pages = {1446},
  year = {1998}
}

@article{wormholes_1988_morris,
  author = {M. S. Morris and K. S. Thorne},
  journal = {American Journal of Physics},
  volume = {56},
  pages = {395},
  year = {1988}
}

@article{casimir1948,
  author = {H. B. G. Casimir},
  title = {On the attraction between two perfectly conducting plates},
  journal = {Proc. Kon. Ned. Akad. Wetenschap},
  volume = {51},
  pages = {793},
  year = {1948}
}

@article{sparnaay1957,
  author = {M. J. Sparnaay},
  title = {Measurements of attractive forces between flat plates},
  journal = {Nature},
  volume = {180},
  pages = {334},
  year = {1957}
}

@article{sparnaay1958,
  author = {M. J. Sparnaay},
  title = {Attractive forces between flat plates},
  journal = {Physica},
  volume = {24},
  pages = {751},
  year = {1958}
}

@article{lamoreaux1997,
  author = {S. K. Lamoreaux},
  title = {Demonstration of the Casimir Force in the 0.6 to 6 $\mu$m Range},
  journal = {Physical Review Letters},
  volume = {78},
  pages = {5--8},
  year = {1997}
}

@article{Garattini2019,
  author = {R. Garattini},
  title = {Casimir Wormholes},
  journal = {European Physical Journal C},
  volume = {79},
  pages = {951},
  year = {2019},
  doi = {10.1140/epjc/s10052-019-7415-1}
}

@article{ford1996,
  author = {L. H. Ford and T. A. Roman},
  title = {Quantum field theory constrains traversable wormhole geometries},
  journal = {Physical Review D},
  volume = {53},
  pages = {5496--5507},
  year = {1996},
  eprint = {gr-qc/9510071},
  archivePrefix = {arXiv}
}

@book{visser1995,
  author = {M. Visser},
  title = {Lorentzian Wormholes: From Einstein to Hawking},
  publisher = {American Institute of Physics},
  address = {New York},
  year = {1995}
}

@article{herrera2024,
  author = {L. Herrera and A. Di Prisco and J. Ospino},
  title = {Ghost stars in general relativity},
  journal = {Symmetry},
  volume = {16},
  pages = {562},
  year = {2024}
}

@article{Varsha:2025oik,
    author = "Varsha, C. S. and Sudharani, L. and Kavya, N. S. and Venkatesha, V.",
    title = "{A novel theoretical approach to dark matter supported Casimir wormhole}",
    doi = "10.1016/j.nuclphysb.2025.116929",
    journal = "Nucl. Phys. B",
    volume = "1017",
    pages = "116929",
    year = "2025"
}

@article{EF1,
  title={Gravitational cracking of stellar models with like-Tolman IV complexity factor},
  author={Andrade, J and Fuenmayor, E and Contreras, E},
  journal={International Journal of Modern Physics D},
  volume={31},
  number={12},
  pages={2250093},
  year={2022},
  publisher={World Scientific}
}

@article{2020contreras,
  title={Contreras, E2020EPJC. 80. 177A. vol. 80},
  author={Abell{\'a}n, G and Torres, V and Fuenmayor, E},
  journal={Eur. Phys. J. C},
  pages={177},
  year={2020}
}

@article{EF2,
  title={2+ 1 Einstein--Klein--Gordon black holes by gravitational decoupling},
  author={Arias, Pio J and Bargue{\~n}o, Pedro and Contreras, Ernesto and Fuenmayor, Ernesto},
  journal={Astronomy},
  volume={1},
  number={1},
  pages={2--14},
  year={2022},
  publisher={MDPI}
}

@article{EF3,
  title={Complexity factor for black holes in the framework of the Newman--Penrose formalism},
  author={Bargue{\~n}o, Pedro and Fuenmayor, Ernesto and Contreras, Ernesto},
  journal={Annals of Physics},
  volume={443},
  pages={169012},
  year={2022},
  publisher={Elsevier}
}

@article{EF0,
  title={Integration of the Lane--Emden equation for relativistic anisotropic polytropes through gravitational decoupling: a novel approach},
  author={Santana, D and Fuenmayor, E and Contreras, Ernesto},
  journal={The European Physical Journal C},
  volume={82},
  number={8},
  pages={703},
  year={2022},
  publisher={Springer}
}

@article{EF4,
  title={Gravitational cracking of general relativistic polytropes: a generalized scheme},
  author={Le{\'o}n, P and Fuenmayor, E and Contreras, E},
  journal={Physical Review D},
  volume={104},
  number={4},
  pages={044053},
  year={2021},
  publisher={APS}
}

@book{Bargueno2023,
    author = {Bargueño, Pedro and Contreras, Ernesto},
    title = "{The Geroch-Held-Penrose Calculus: Fundamentals and Applications}",
    publisher = {Springer},
    series = {Springer Briefs in Physics},
    place = {New York},
    year = {2023}
}

@article{Senovilla1998,
  title = {Singularity Theorems and Their Consequences},
  volume = {30},
  ISSN = {1572-9532},
  url = {http://dx.doi.org/10.1023/A:1018801101244},
  DOI = {10.1023/a:1018801101244},
  number = {5},
  journal = {General Relativity and Gravitation},
  publisher = {Springer Science and Business Media LLC},
  author = {Senovilla,  José M. M.},
  year = {1998},
  month = may,
  pages = {701–848}
}

@article{Misner1964,
  title = {Relativistic Equations for Adiabatic,  Spherically Symmetric Gravitational Collapse},
  volume = {136},
  ISSN = {0031-899X},
  url = {http://dx.doi.org/10.1103/PhysRev.136.B571},
  DOI = {10.1103/physrev.136.b571},
  number = {2B},
  journal = {Physical Review},
  publisher = {American Physical Society (APS)},
  author = {Misner,  Charles W. and Sharp,  David H.},
  year = {1964},
  month = oct,
  pages = {B571–B576}
}

@article{Hawking1968,
  title = {Gravitational Radiation in an Expanding Universe},
  volume = {9},
  ISSN = {1089-7658},
  url = {http://dx.doi.org/10.1063/1.1664615},
  DOI = {10.1063/1.1664615},
  number = {4},
  journal = {Journal of Mathematical Physics},
  publisher = {AIP Publishing},
  author = {Hawking,  S. W.},
  year = {1968},
  month = apr,
  pages = {598–604}
}

@book{Penrose1984,
    author = {Penrose, R. and Rindler, W.},
    title = "{Spinors and Space-Time. Volume 1: Two-Spinor Calculus and Relativistic Fields}",
    publisher = {Cambridge University Press},
    series = {Cambridge Monographs on Mathematical Physics},
    place = {Cambridge},
    year = {1984}
}

@article{Hayward1994,
  title = {Quasilocal gravitational energy},
  author = {Hayward, Sean A.},
  journal = {Phys. Rev. D},
  volume = {49},
  issue = {2},
  pages = {831--839},
  numpages = {0},
  year = {1994},
  month = {Jan},
  publisher = {American Physical Society},
  doi = {10.1103/PhysRevD.49.831},
  url = {https://link.aps.org/doi/10.1103/PhysRevD.49.831}
}

@article{LHnegative,
  title={Negative energy density and classical electron models},
  author={Herrera, L and Verela, V},
  journal={Physics Letters A},
  volume={189},
  number={1-2},
  pages={11--14},
  year={1994},
  publisher={Elsevier}
}

@article{Najernegative,
  title={On negative mass cosmology in General Relativity},
  author={N{\'a}jera, Sebasti{\'a}n and Gamboa, Aldo and Aguilar-Nieto, Alejandro and Escamilla-Rivera, Celia},
  journal={Astronomy \& Astrophysics},
  volume={651},
  pages={L13},
  year={2021},
  publisher={EDP Sciences}
}

@article{Farnesnegative,
  title={A unifying theory of dark energy and dark matter: Negative masses and matter creation within a modified $\Lambda$CDM framework},
  author={Farnes, Jamie Stephen},
  journal={Astronomy \& Astrophysics},
  volume={620},
  pages={A92},
  year={2018},
  publisher={EDP Sciences}
}

@article{Naseerpossible,
  title={Possible existence of ghost stars in the context of electromagnetic field},
  author={Naseer, Tayyab and Hassan, K and Sharif, M},
  journal={Chinese Journal of Physics},
  volume={94},
  pages={594--608},
  year={2025},
  publisher={Elsevier}
}

@article{Herreraevolution,
  title={Evolution of self-gravitating fluid spheres involving ghost stars},
  author={Herrera, Luis and Di Prisco, Alicia and Ospino, Justo},
  journal={Symmetry},
  volume={16},
  number={11},
  pages={1422},
  year={2024},
  publisher={MDPI}
}

@article{Herrerabirth,
  title={The birth of a ghost star},
  author={Herrera, Luis and Di Prisco, Alicia and Ospino, Justo},
  journal={Entropy},
  volume={27},
  number={4},
  pages={412},
  year={2025},
  publisher={MDPI}
}

@article{Herreraaxially,
  title={Axially symmetric ghost stars},
  author={Herrera, L and Hern{\'a}ndez--Pastora, JL and Ospino, J and Di Prisco, A},
  journal={The European Physical Journal C},
  volume={85},
  number={6},
  pages={618},
  year={2025},
  publisher={Springer}
}

@article{herrera2001conformally,
  title={Conformally flat anisotropic spheres in general relativity},
  author={Herrera, L and Prisco, A Di and Ospino, J and Fuenmayor, E},
  journal={Journal of Mathematical Physics},
  volume={42},
  number={5},
  pages={2129--2143},
  year={2001},
  publisher={American Institute of Physics}
}

@article{herrera2018new,
  title={New definition of complexity for self-gravitating fluid distributions: The spherically symmetric, static case},
  author={Herrera, L},
  journal={Physical Review D},
  volume={97},
  number={4},
  pages={044010},
  year={2018},
  publisher={APS}
}

@article{arias2022anisotropic,
  title={Anisotropic star models in the context of vanishing complexity},
  author={Arias, C and Contreras, E and Fuenmayor, E and Ramos, A},
  journal={Annals of Physics},
  volume={436},
  pages={168671},
  year={2022},
  publisher={Elsevier}
}

@article{herrera2018definition,
  title={Definition of complexity for dynamical spherically symmetric dissipative self-gravitating fluid distributions},
  author={Herrera, L and Di Prisco, A and Ospino, J},
  journal={Physical Review D},
  volume={98},
  number={10},
  pages={104059},
  year={2018},
  publisher={APS}
}

@misc{herrera2021complexity,
  title={Complexity of self-gravitating systems},
  author={Herrera, Luis},
  journal={Entropy},
  volume={23},
  number={7},
  pages={802},
  year={2021},
  publisher={MDPI}
}

@article{turimov2025exact,
  title={Exact charged traversable wormhole solution},
  author={Turimov, Bobur and Abdujabbarov, Ahmadjon and Ahmedov, Bobomurat and Stuchl{\'\i}k, Zden{\v{e}}k},
  journal={Physics Letters B},
  pages={139800},
  year={2025},
  publisher={Elsevier}
}

@article{canate2024traversable,
  title={Traversable wormholes with electric and magnetic charges in general relativity theory},
  author={Ca{\~n}ate, Pedro},
  journal={Physical Review D},
  volume={110},
  number={8},
  pages={084030},
  year={2024},
  publisher={APS}
}

@article{neto2023shadow,
  title={The shadow of charged traversable wormholes},
  author={Neto, M{\'a}rio Raia and P{\'e}rez, Daniela and Pelle, Joaquin},
  journal={International Journal of Modern Physics D},
  volume={32},
  number={02},
  pages={2250137},
  year={2023},
  publisher={World Scientific}
}

@article{avalos2025hyperbolic,
  title={Hyperbolic Casimir-like wormhole},
  author={Avalos, Roberto and Brito, D and Fuenmayor, E and Contreras, E},
  journal={The European Physical Journal C},
  volume={85},
  number={7},
  pages={1--14},
  year={2025},
  publisher={Springer}
}

@article{bhattacharya2023complexity,
  title={Complexity factor parameterization for traversable wormholes},
  author={Bhattacharya, Subhra and Nalui, Subhasis},
  journal={Journal of Mathematical Physics},
  volume={64},
  number={5},
  year={2023},
  publisher={AIP Publishing}
}

@article{ziaie2024casimir,
  title={Casimir wormholes in Brans--Dicke theory},
  author={Ziaie, Amir Hadi and Mehdizadeh, Mohammad Reza},
  journal={Classical and Quantum Gravity},
  volume={41},
  number={14},
  pages={145001},
  year={2024},
  publisher={IOP Publishing}
}

@article{sokoliuk2022probing,
  title={Probing the existence of the ZTF Casimir wormholes in the framework of f (R) gravity},
  author={Sokoliuk, Oleksii and Baransky, Alexander and Sahoo, PK},
  journal={Nuclear Physics B},
  volume={980},
  pages={115845},
  year={2022},
  publisher={Elsevier}
}

@article{varsha2025novel,
  title={A novel theoretical approach to dark matter supported Casimir wormhole},
  author={Varsha, CS and Sudharani, L and Kavya, NS and Venkatesha, V},
  journal={Nuclear Physics B},
  pages={116929},
  year={2025},
  publisher={Elsevier}
}

@article{avalos2022traversable,
  title={Traversable wormholes with like-{C}asimir complexity supported with arbitrarily small amount of exotic matter},
  author={Avalos, R and Fuenmayor, E and Contreras, E},
  journal={The European Physical Journal C},
  volume={82},
  number={5},
  pages={420},
  year={2022},
  publisher={Springer}
}

@book{Darmois1927,
  author    = {Georges Darmois},
  title     = {Les équations de la gravitation einsteinienne},
  series    = {Mémorial des sciences mathématiques},
  number    = {25},
  year      = {1927},
  pages     = {58},
  publisher = {Gauthier-Villars},
  url       = {https://www.numdam.org/item/MSM_1927__25__1_0/}
}

@article{Israel1966,
  title = {Singular hypersurfaces and thin shells in general relativity},
  volume = {44},
  ISSN = {1826-9877},
  url = {http://dx.doi.org/10.1007/BF02710419},
  DOI = {10.1007/bf02710419},
  number = {1},
  journal = {Il Nuovo Cimento B Series 10},
  publisher = {Springer Science and Business Media LLC},
  author = {Israel,  W.},
  year = {1966},
  month = jul,
  pages = {1–14}
}

@article{Mena2008,
author = {Filipe C. Mena and Jos{\'e} Nat{\'a}rio and Paul Tod},
title = {{Gravitational collapse to toroidal and higher genus asymptotically AdS black holes}},
volume = {12},
journal = {Advances in Theoretical and Mathematical Physics},
number = {5},
publisher = {International Press of Boston},
pages = {1163 -- 1181},
year = {2008},
}

\end{document}